\documentstyle[11pt,aasms4]{article}
\newcommand{\eti}{ et al. \/}
\newcommand{\sfr}{star formation rate }
\newcommand{\ci}{\'{\i}}
\lefthead{Tissera, P.B.}
\righthead{Star Formation}

\begin{document}

\title{Analysis of Star Formation  in  Galaxy-like Objects}

\author{Patricia B. Tissera \altaffilmark{1}}
\altaffiltext{1} { I.A.F.E., Casilla de Correos  67, Suc. 28, Buenos
Aires (1428), Argentina. E-mail: patricia@iafe.uba.ar
}

\begin{abstract}

Using cosmological hydrodynamical simulations,
 we investigate the effects of hierarchical 
aggregation on the triggering of star formation in galactic-like
objects. We include a simple star formation model
to transform the cold gas in dense regions into stars. Simulations
with different parameters have been performed in order to quantify 
 the dependence of the results on the parameters.
We then resort to stellar population synthesis models to trace
the color evolution of each object  with red-shift and  in 
relation to their merger histories.
We find that, in a hierarchical clustering scenario, the process of
assembling of the structure is one natural mechanism that
may trigger star formation. 
 The resulting
star formation  rate history for each individual galactic object is
composed of a continuous one ($\leq 3 \rm{M_{\odot}/yr}$) and
a series of star bursts. 
 We  find that even the accretion of a small satellite
can be correlated with a stellar burst.
Massive mergers are found to be more efficient at transforming
 gas into stars

\end{abstract}


\section{Introduction}

Within the last few years,
 it has been possible to start drawing an observational
 picture of the formation and evolution of galaxies 
(e.g. \cite{fuk96}) based upon
 an unprecedented amount of
 results on the astrophysical properties  of galactic objects
up to $z\approx 5$ at
ultraviolet, optical and near IR wavelengths (e.g. \cite{sh92}; 
 \cite{mad95}; \cite{lil95}; \cite{cow96};  \cite{ell96};
\cite{ste98}).
These data provide observational constraints on the history of
structure formation in the Universe. In particular, there has
been a significant progress in describing the
cosmic history of star formation and metal enrichment (e.g.
 \cite{mad96}; \cite{low97}).
Hopefully these data together with previous and future results will
contribute to the understanding of the  process of star formation.
Although there are several works on this subject in the scientific literature
(e.g. \cite{ken96}; \cite{fer97} and references therein; \cite{ken98}), 
the detailed physical mechanism is still under study since
several aspects remain poorly understood.
One open question is regarding the factors that trigger and control 
the star formation activity in galaxies of different morphologies.

In a hierarchical clustering model where structure forms
in a bottom-up way, a typical galactic halo is the result of
the mergers of smaller substructures. 
The  halo merger frequency  and the typical mass involved in one event vary
depending on the environment and the cosmological model assumed,
among other factors.
The astrophysical properties of the galactic objects
are determined by all these processes and by the hydro-dynamical 
evolution of the baryons in a complex way.
The conditions which may
trigger and control the transformation of cold gas in dense regions
into stars are not yet well understood and may involve  different
processes such as supernova energy injection, mergers, disk instabilities
and tidal fields.
Of particular interest  is the possible triggering of star formation
by the  merging of substructures as a galactic object is assembled
in a hierarchical clustering scenario. This fact would imply  the existence 
of a correlation 
between mergers with satellites and enhancements of 
the star formation activity. 
 Hereafter, we will define a merger
 as the complete
process since the time two baryonic objects are identified to  share the
same dark matter halo for the first time to their actual fusion. 
And whenever tidal fields are mentioned, they
are those originated during the mergers. The study of any  interactions that
do not result in the actual fusion of the structures is not considered 
 in this paper.

There has been also much work on the link between
the high star formation rates in ultra-luminous IRAS galaxies (ULIRGs) and their
environments, although a consensus on how the galaxy environment influences
its \sfr has yet to be reached. Lawrence et al. (1989) find that most ($\sim 70\%$)
of the ULIRGS they studied appeared to have close companions or be morphologically
disturbed (\cite{luc97}; see also \cite{sm96}),
 whilst a significant minority appear to be isolated.
Other authors such as Clements et al. (1996) and Sanders et al. 
(1988) estimate that $\sim 99\%$
of ULIRGs are in interacting systems. It is clear that much of the discrepancy is due to
the difficulty in classifying quantitatively what is meant by `interacting'.
Despite this reliance on a rather subjective classification of the data,
there seems to be a  trend in increasing likelihood of interaction and increasing starburst
luminosity in IRAS selected galaxies. 
All the above studies of the environments of star-bursting galaxies assume
that both, the environmental trigger and the starburst episode, are contemporaneous.
As Joseph and Wright  (1985) point out, this might not be true if 
we use tidal tails
as the evidence for an interaction as, depending on the stellar initial mass
function (IMF), the tidal tail might last
longer than the starburst it triggers.
Therefore it is still unclear as to whether an interaction/merger
is {\em required} in order for a starburst to take place and also whether they happen
simultaneously.

On the other hand, recent observations (mainly from the Hubble Deep Field)
show an increasing number of irregular/interacting or morphologically
disturbed objects with look-back time, which also exhibit strong
star formation activity (e.g. \cite{gla95}; \cite{ell96};
\cite{lil97}; \cite{low97}; \cite{guz97}; \cite{dri98}).
It is not yet clear at what extent these data are fully consistent with
predictions of models based on hierarchical clustering. Nevertheless all
of them agree in supporting a scenario where interactions/mergers
and star formation activity increase with red-shift (e.g. \cite{bou97};
\cite{dri98}; \cite{bri98}).

From a theoretical point of view,
semi-analytic models have been quite successful in formulating a
picture of galaxy formation and evolution, although they have
to resort to  global
recipes in order to take into account complex effects (\cite{kau93};
\cite{bau96}).
In this regard, Lacey \& Silk (1991) have also used tidal interactions
between galaxy-sized objects for
controlling the star formation processes in galactic objects,
 although they did not consider  
mergers of substructure 
 as a possible cause. Baugh et al. (1996) formulate a semi-analytic model where the morphology of
a galaxy  is determined by its history of major mergers which
trigger violent star formation. No predictions for the strength,
duration and frequency of these stellar bursts could be done with this model.

Self-consistent numerical simulations have proved a powerful
tool for the study
of the formation of galaxies in a cosmological framework (\cite{cen98}; \cite{her96}; \cite{the98}; \cite{tis97};
\cite{pea99}; \cite{kat99}).
 They have
the advantage over semi-analytic models of being able to
provide a consistent description of the evolution of the structure
in the non-linear regime. As a consequence, physical processes related to
the evolution of the dissipative component can be included and modeled upon a more
physical basis. Among these processes, star formation mechanisms
are highly important because of their outstanding role
in the formation of structure on galactic scales.
 However, the treatment of processes related
with star formation is at its beginnings, since numerical problems,
added to the lack of a fully theoretical understanding, make 
its implementation difficult.
Several authors have analyzed numerical models of the formation of 
individual galaxies
including simple schemes to transform the cold dense
gas into stars (e.g. \cite{kat92};
\cite{nw93}, \cite{ger97}). 

Of most relevance to the analysis carried out in this paper are the
results of
Barnes \& Hernquist (1991, 1996) and Mihos \& Hernquist (1994, 1996).
 These authors  modeled 
 mergers of two disk/halo galaxies using hydro-dynamical codes.
 They found that a merger with a satellite can induce  the 
formation of a bar along
which the gas is compressed and 
shocked, loosing angular momentum. This process would
trigger a gas inflow which could fuel star formation
activity at the center.
These models illustrate one way in which  tidal
fields can produce gas inflows in strongly interacting and merging galaxies.
Mihos and Hernquist (1996) also showed that the internal structure of a parent disk-like
galaxy is relevant to regulate the rate of this gas inflow.
Hernquist (1989 a,b) showed that even the accretion of low-mass
satellites by disks can result in an inward gas flow.
In all cases, by applying the Schmidt law, an enhancement of the
gas density directly implies an increase of the star 
formation activity.

 With the models described in this work, we intend to
establish the relevant characteristics of the interplay between
hierarchical aggregation  and star formation, and   to assess if the 
outcomings
are consistent with observations of galaxies undergoing  star formation
activity in a cosmological framework.
Note, however, that we intend to analyze the star formation
process in {\it normal} field galactic-like objects. We will refer
 to  mergers  that arise as a consequence of
their formation and evolution in a hierarchical cosmological scenario.
In hierarchical clustering scenarios, galactic objects build up through the
aggregation of substructure and may suffer major and minor encounters throughout their life.
Their star formation histories may be affected by these mergers and, as already
pointed out by several authors (e.g., \cite{mh96}, \cite{bh96}; \cite{bau96}), they may contribute to its triggering.
As an attempt to assess the relation between this hierarchical built-up of
the structure and the
  star formation process, we analyze the evolutionary history of each halo in our
simulations, looking for  possible correlations between
star formation enhancements and mergers. 
The main difference between previous works and this one, is that, in this case, every merger
event arises naturally, in consistency with a cosmological model.
In fully consistent cosmological simulations, the distribution
of merger parameters, such as the orbit characteristics, the 
orbital energy and angular momentum, the masses of the
virial halos and baryonic clumps involved in the merger event,
and the spin, internal structure and relative orientation of the baryonic
clumps that are about to merge, among others, arise naturally
at {\it each epoch} as a consequence of the initial spectrum
of the density fluctuation field, its normalization, and the cosmological
models and its parameters. In controlled type mergers, they 
are set by hand.
Moreover, in fully-consistent cosmological simulations,
the effects of diffuse gas accretion and of interactions
with small satellites on the assembly of a galactic 
object are also accounted for.
  
We will also intend to know at what extent can a simple star
formation (SF) model follow the SF history of a galaxy-like object, if those
SF histories are consistent with observations and how sensitive they
are to the free parameters of the model.
We will also use the technique  described by Tissera, Lambas \& Abadi
(1997) to assign luminosities at different wavelengths  to
the simulated galactic objects. This implementation
allows us  to follow the evolution of their colors as
a function of $z$ and in relation to their  evolutionary
history.

This paper is organized as follows. Section 2 describes the main
characteristics and parameters of the simulations. Section 3 analyses
the results and compares them with recent observational data.
In Section 4  we investigate the effects of mergers on 
the color distributions of the galaxies formed,
and Section 5 outlines the  results.
\section{Simulations}
 The simulations analyzed
follow the evolution of a typical region of the Universe
 using a version of  AP3M+SPH code (\cite{tis97}).
We carried out three simulations (hereafter referred as S.1, S.2 and S.3).
S.1 and S.2 share the same
initial conditions for the distributions of gas and dark matter particles, 
whilst S.3 has different ones.
Simulations S.1 and S.2 are identical except for the star formation
efficiency, so any difference between them is due to 
the SF process and the fact that stars behave as collision-less particles.

The initial conditions are set up by using ACTION (for S.1 and S.2)
 and COSMICS (for S.3),
and are consistent
with a Cold Dark Matter (CDM) spectrum with $\Omega =1$,
$\Lambda =0$, 
$\Omega_{\rm b}= 0.1$, and 
$\sigma_{8}=0.4$  for S.1, S.2 and $\sigma_{8}=0.67$ for S.3.
We used  $N = 262144$
particles 
(${\rm N_{dark}= 259520}$ and ${\rm  N_{bar}=26214}$) 
in a
comoving box of length $L=5 h^{-1} $ Mpc ($H_{0}=100 h^{-1}\ {\rm{ km \
s^{-1}\ Mpc ^{-1}}}$, $h=0.5$).
Note that dark matter and baryonic particles have the same mass 
(${\rm M_{part}=2.6 \times 10^{8} M_{\odot}}$).
 The gravitational  softening used in these simulations is 3 kpc, and
the smaller smoothing length allowed is 1.5 kpc.
The time-steps of integration used are $\Delta t=1.4 \times 10^7$ yr for S.1 and S.2, and $\Delta t =
1.2 \times 10^7$ yr for S.3.
These simulations have proved to be adequate to study processes
related with formation of galaxies in a fully
cosmological framework (e.g., \cite{tis97}; \cite{td98}; \cite{dom98};
\cite{tis99}) despite they have lower
hydro-dynamical resolution when compared to prepared ones 
(\cite{ns97}).

These simulations include star formation  according to the
 algorithm described by \cite{tis97}.
 The gas cools due to radiative cooling.
We use the approximation for the cooling function
 given by Dalgarno and McCray (1972).
Gas particles are transformed into stars if they are cold 
($T_* \leq 3\times 10^{4}$ K
for S.1 and S.2, and $T_* \leq 10^{4}$ K for S.3),
dense ($\rho > \rho_{\rm crit}\approx 7\times 10^{-26}
\rm{gr \ cm^{-3}}$) and
satisfy the Jean's instability criterium.
When a gas particle satisfies these conditions, it is
transformed into a star particle after a time interval ($\tau$)\footnote{
This time interval is the time estimated from equation (1) over which
99 $\%$ of the gas mass in a particle  is expected to be transformed into stars (\cite{nw93}).
 It is
estimated as: $\tau ={\rm ln} 0.01 t_{*}/c$.}
over which its gas mass is being converted into stars according to

\begin{equation}
 \frac{d\rho_{\rm star}}{dt}=-c \frac{\rho_{\rm gas}}{t_{*}}
\end{equation}

  where $c$ is the
star formation efficiency ($c=0.01,0.1, 0.01$ for
S.1, S.2 and S.3, respectively) and $t_{*}$ is a characteristic time-scale
assumed to be equal to the dynamical time of the particle 
($t_* = t_{\rm dyn} = (3 \pi /(16G
\rho_{\rm gas}))^{1/2}$).

 The simulations  have different
star formation efficiency parameters, $c$. The total number of stars formed
will depend on  the values of $c$ and $T_*$,  which can be adjusted in order to reproduce
observations.
Simulations S.1, S.2 and S.3 have transformed $12 \%$, $28 \%$
and $7.5 \%$, respectively,
of their total baryonic mass into stars at
$z=0$. 
Because of the higher value of $\sigma_{8}$ used in S.3,
halos collapse earlier and reach higher
densities sooner. As a consequence, stars form from higher $z$
 depleting quicker the gas in condition
of forming stars. Hence, the SF history depends also on the 
normalization of the power spectrum (\cite{bau96}).
Different values of critical temperature $T_*$ have been used.
Simulation S.3 has a lower value which helps to produce less
stars than in S.1.
The total stellar masses at $z=0$ imply a stellar density parameter 
$\Omega_{\rm s}$
 greater than the observed ones: $0.005 < \Omega_{\rm s} h^{2} <0.009$ (\cite{mad98}),
although the latter are subject to  serious  uncertainties such as 
dust effects which can lead to important underestimations.
Recent observations in the mid and far infrared
suggest higher values (\cite{flo98}).

Note also that we have adopted $\Omega_{\rm b} = 0.10$, a value
that could be considered somewhat high. However, recent measurements
of the deuterium abundance in clouds of hydrogen at high
red-shift (\cite{bt98a};\cite{bt98b}), if correct,
allow to constrain the baryon fraction to a precision of
$10\%$, $\Omega_{\rm b} = 0.08 \pm 0.008 $ (for $h=0.5$, as we
have assumed), close to the value we have used.  
 

Feedback effects and metallicity enrichment  by supernova explosions
have not been included in these simulations. Feedback processes are
believed to play a key role in helping to set a self-regulated star formation
regimen (\cite{sil97}). But its modeling in hydro-dynamical simulations
is still quite controversial (\cite{kat92}; 
\cite{nw94}; \cite{me95};
\cite{yep97}).
Since the SF process is not completely understood, it is wise to analyze
these two effects in separate steps.

One shortcoming of numerical simulations is that numerical resolution decreases
with look-back time. The higher the red-shift, the smaller the objects and so,
the smaller the number of particles used to resolve them. This is an
ubiquitous problem
in all numerical simulations and a very large number of particles
would be required to improve it. Since this is impossible to
accomplish at present, results should always be considered with caution.
In order to minimize this problem, we will study the evolutionary
history of the larger objects in our simulations (see Section 3).
Note, however, that we use fully self-consistent cosmological simulations, hence, 
the hierarchical
evolution of a galactic halo is very well represented and so are
mergers and 
 the effects of tidal fields generated by the nearby structure.

\section{Analysis}

\subsection{Global Star Formation}

Recent observations of objects at different $z$ have provided with information
about the  SF history of the Universe (\cite{sp98} an references therein). Although many uncertainties such as
the initial mass function, reddening by internal
and Ly$\alpha$ cloud absorptions and the fact that UV luminosity
traces mainly the formation of massive stars, among others,
  affect these results, it is possible to envisage a global SF trend.
In this sense, this observational relation can be a useful tool to assess how the
SF proceeds within a simulated box. It has to stressed that we do not intend
to explain what is observed but to use observations
 as a global constrain to qualitatively compare
the effects that the different SF parameters may have on the SF history of the
simulated box.

We estimate the global SFR by reckoning the stellar mass formed
in the simulated box at each $z$, and smoothing it overtime
in order to diminish the noise introduced by the discreteness of the SF
process.
In order to compare these  mean star formation rate $<SFR>$
 histories
with observational results,
we calculate the cosmic  star formation rate density:
$\rho_{\rm SFR} = <SFR>/ {\rm V }$, at each time-step of integration
in each simulation.  V is the comoving
volume of the simulated boxes at each corresponding $z$.

In Figure 1, we plot $\rho_{SFR}$ for
simulations S.1, S.2 and  simulation S.3,
 and
  include recent observational results.
As can be seen from this figure, the simulated $\rho_{SFR}$ are quite
different. None of them have a peak at $z\approx 1.5$
as that claimed by Madau et al. (1996), but they
are within the observed range.

Simulations S.1 and S.2, because of the combined effects of the
lower normalization parameter $\sigma_8$ and SF ones, they start forming
stars later on. The only difference between S.1 and S.2 is the value of $c$.
Therefore, it is direct that a change in the star formation efficiency introduces a delay in process
and decreases the overall star formation rates.
To fill up the gap between $z\approx 2$ and the last point measured by
Madau \eti (1996) at $z=5.5$, it can be estimated that approximately 20 gas particles should have been transformed into stars at a constant rate of 45 $ {\rm M_{\odot}/yr}$.
This number represents less than a $10\%$ of the total stellar mass formed
in S.1 and S.2. Whence we can conclude that the results for $z<2$ will
not be strongly affected by the necessary changes in the SF process
in order to have a complete SFR history at higher $z$.

Simulation S.3  starts forming stars at larger $z$  due
to the higher normalization parameter $\sigma_{8}$ adopted. But as a consequence
of the low $c$ and $T_*$ values used, the SF rates are lower than
those in the other two simulations.
Globally, the $<SFR>$ in S.3 shows a  similar trend
to observations. In this simulation, the total amount of stars formed at $z>2$
is $30\%$ of
the total stellar mass at $z=0$, while $55\%$ was formed at $1<z<2$.
Note that the peak of SF is actually at earlier times, $z\approx 3.5$, and 
that it decreases slowly to smaller $z$ in accordance with recent results from
Cowie, Songaila and Barger (1999).

It has to be mentioned that numerical resolution affects SF more strongly at higher $z$
that at lower ones, and, that the inclusion of feedback mechanisms could have a non-negligible
impact of the $\rho_{\rm SFR}$. However, current numerical models that include SN feedback are still
limited by numerical and theoretical problems.

To sum up, the normalization of the power spectrum, the
star formation efficiency and the minimum $T_*$
adopted in the SF model, affect the global SFR history
of the simulated volumes and consequently, the SFR of
each individual galaxy-like object.
The value of $T_*$ is determined, in part by the available
cooling functions, and indirectly, by numerical
resolution of the gas component.
The normalization parameter $\sigma_8$
is now defined within a narrow range depending on
the adopted cosmology. However, many questions
remain to be answered about the bias and its dependence
on scale and red-shift. The value of $c$
is the actual free parameter of the SF model.
In the following sections, we use
these experiments (S.1, S.2 and S.3) 
to analyze the SF process within 
each galaxy-like object in relation to their
merger histories, and assess how these parameters affect the individual SF histories.

\subsection{Star Formation History and Mergers}

In this paper, we restrict the analysis to typical field  
galactic-like objects (GLOs). At $z=0$, GLOs are
identified at their virial radius i.e., the  radius
for which the density contrast is  estimated to be
$\delta \rho /\rho \approx 200$ (\cite{wf91}).
We reject those GLOs with a comparable companion within
two virial radii at $z=0$. In this way, we avoid complications due to
 tidal fields originated by the presence of several companions and/or
the underlying over-density, focusing on the effects produced
by the assembly of each individual object through hierarchical
growth. In particular, in S.3 some objects have been
discarded since they belong to groups.
Each GLO is composed of a dark matter halo and a baryonic
component in the form of gas and/or stars.
We will only analyze in detail objects resolved with more
than 250 baryonic particles within their virial radius at $z=0$.
 Table 1 gives the total
dark  matter(${\rm N_{\rm dark}}$) and baryonic (${\rm N_{\rm bar}}$) particles
within the virial radius for each GLO at $z=0$.
 The SF algorithm used is very effective at forming stars at
the dense cores of the galactic-like objects, so it is easy to isolate star 
particle clumps.

We have named GLOs in S.1 and S.2  using the same label code
as the one chosen in Tissera \& Dom\ci nguez-Tenreiro (1998) to identify 
halos. Simulations S.1 and S.2 correspond to their simulations I.2 and I.3.
The main  baryonic clumps in GLOs in S.1 that resemble
a disk-like structure (DLO)  have been studied from a dynamical point
of view by  Dom\ci nguez-Tenreiro et al. (1998). GLOs 3, 4, 5
and 6 in Table 1 host DLOs 1, 2, 4 and 3 in 
their Table 1, respectively.

 We then follow back  the evolution of the matter inside
their virial radius as a  function
of the look-back time for the available outputs of the simulations.
We construct the merger trees of each GLO identified at $z=0$ in 
the three simulations by recursively tracing back in time the objects
which contain particles that end up in the final GLO.
In this way, we individualize the progenitors and the satellites with
which they merged at all outputs of the simulations (every
100 time-steps for S.1 and S.2, and 20 time-steps for S.3).
All objects, progenitors and satellites, are identified at their virial
radius at the corresponding $z$. 
The set of objects identified in this way gives a complete record of
the merger history of each GLO.
We will assume that the progenitor clump or parent galaxy of a GLO
at $z=0$ is the most massive
object identified at $z > 3$ from its merger tree. 
A merger will be defined 
 as the complete
process since the time two baryonic objects are identified to  share the
same dark matter halo for the first time to their actual fusion. 
We will not distinguish  between major and minor mergers.
Instead, a merger will be counted each time the progenitor fusions with  a 
satellite
with more than 10 \% of its virial 
mass at that time of the merger.
 Otherwise, (i.e. less than 10 \%) it will be considered accretion
or infall.
We can, then, follow the evolution with look-back time of  each dark matter halo and its baryonic
main clump.
Some small satellites may have been formed and accreted between outputs.
In this case, they would be missed by our analysis. However, this situation
would happened only for very small objects that can be, anyway, counted as
accretion. Note that we keep track of evolution
of all smaller  substructures that merge with the progenitor, but we do not look at them in detail as only their virial mass and gas content
at the time of the merger are required for this analysis.


 We  estimate the star formation history of each GLO by
reckoning the stellar mass formed in its progenitor
objects at each $z$ and then, smoothing these distributions
over time.
The reason for adopting this procedure is that the 
SF model used in these simulations transforms a gas particle
into a star one at once, after a time delay $\delta t$ over
which the gas is supposed to be transformed into stars as explained in
Section 2. A typical value for this time delay  is $<\delta t> \approx 20 \
\Delta t$, where $\Delta t$ is the integration time-step of the
simulations
Then  the  $SFR$ have been smoothed  by
binning these distributions in time-bins of 20 points centered
at the formation time of each star particle and averaging the
stellar mass formed within each time-bin.

In Figure 2 we show, as an example, the star formation history
from $z=1$ of the  galactic objects 1, 2, 3 and 4 
 in simulation S.1.
We have plotted the $<SFR>$ (${\rm M_{\odot}/yr}$)
in the progenitor object 
 versus look-back time ($\tau(z)=1.-(1+z)^{-2/3}$ for $\Omega=1$ ).
The times at which the satellite enters
the virial radius of the progenitor has been indicated with
an arrow pointing up, while  the actual
fusion of the baryonic cores has been indicated with
an arrow pointing down. 
These are all the merger events in which the GLOs are involved
in the  range depicted in the figure.
As can be seen in Figure 2,
in all cases there is an increase of SF related to a merger with a satellite
of more than 10 \% the progenitor mass. This situation is 
common to all GLOs in the other three simulations.
From this figure it can be also seen
that  when a
satellite enters the virial region  of the progenitor, there is a delay
in the fusion of the gaseous cores (\cite{nav95}).
During this
interval, the objects orbit around each other and
are under the effects of
strong tidal fields. According to the analysis of some authors
 (e.g., \cite{her89a}; \cite{her89b}; \cite{bar88}; \cite{dom98}),
the interactions and fusions with satellites may supply  gas
to the central region of the parent galaxy fueling a burst of star formation.

Recall that among the different processes that can  affect the 
 SF history of the GLOs in hierarchical clustering scenarios
 (such as the assembly of the main object at high $z$,
 the merger
of the progenitor with other clumps, gas compression  
 as it cools and collapses inside dark matter halos during the quiescent phases
of the assembly of the GLOs,
 and the interactions with neighboring structures that do not end up
in actual fusions),
in this work, we refer only to mergers (as defined in the Introduction).
 Consequently, we only study
those well-defined  peaks that are located within merger events (i.e., those
within arrows in Figure 2). So we have not analyzed  the first
SF peaks that can be related with the assembly of the progenitor object
(i.e., in Figure 2   peaks at $\tau(z)\approx 0.32$ in GLO 1 and 4). 
These peaks occur at a $z$ where the  virial mass of the progenitor is always  less than $\leq 20\%$ of the final GLO virial mass, and they
could be strongly affected by  numerical
resolution.


During the merger process (i.e. from the time the satellite enters the virial radius
of the progenitor until one single baryonic clump forms) the progenitor
and its satellite continue accreting gas.
 In our simulations, 
this fraction is  important since,
in same cases, it equals the amount of new stars formed.
For example, GLO 2 in S1 transformed $30\%$ of its original gas
into stars during a merger with an object with a mass of $40\%$ the progenitor mass.
During that process,
the amount of gas accreted was $29\%$ of the  gas mass of the final objects,
and the ratio
between the new stars and the
old ones was $0.82$. The same object in S.2 transformed $83\%$ of its
initial total gas into stars, accreted $30\%$ of
the remnant gas and the burst resulted in a $25\%$ increase of
stellar mass.

\subsection{Star Formation Peaks}
Because the star formation algorithm used in this
work, in practice, transforms at once (after satisfying all 
the requirements
mentioned in Section 2)  a gas particle into  a star one, the overall
star formation history of a galaxy-like object is discrete and quite noisy.
Although it is clear when there is a peak in the star formation history, 
this noise  makes it difficult to isolate the stars formed in
a single burst, and consequently, to classify
the strength of the peak of  new  stars.
In order to do so more rigorously, we took the following
steps. We estimate
the overall minimum star formation rate in a GLO,
 $\delta_{\rm min}$, at any red-shift.
We, then, subtract
a factor $f$ of this minimum from the total star formation rate
 history, so the peaks are clearly
identified as the values with a signal larger than 
a threshold, $\sigma_{\rm min}=f\times\delta_{\rm min}$.
We tried different values of  $f$, choosing $f=3$ since this is the
minimum  one that
allows us to individualize peaks
 in all GLOs in all simulations.
Values below $\sigma_{\rm min}$ are considered 'ambient star formation rate' (ASFR).
This ASFR can be explained as being driven
by the increase of cold dense gas as the result of 
the cooling and collapse  of baryons on to the potential well of the halo.
We estimate that the  ASFRs  take  values of  
$\leq 3 {\rm M_{\odot}\ yr^{-1}}$. The mean values for $<\delta_{\rm min}>$
are 0.90, 1.45, 0.89 $\rm M_{\odot}/yr$ for S.1, S.2 and S.3, respectively.
The larger mean value measured for S.2  reflects the fact that
the rate of star formation is always higher in this simulations
 (since  $c$ is higher) than
in the other ones.
In Figure 2, the horizontal solid lines represent  $\sigma_{\rm min}$ for each
GLO.
 From this figure we see that the total star formation rate histories are
composed of this approximately constant ambient star formation
rate over which stellar bursts are superposed.

 \begin{deluxetable}{crrrrrrrr}
\footnotesize
\tablecaption{Main characteristics of star  bursts.  \label{tbl-2}}
\tablewidth{0pt}
\tablehead{
\colhead{S} & \colhead{$GLO$} &   \colhead{ $\rm N_{dark}$}&
 \colhead{$\rm N_{bar}$}& \colhead{$\sigma_{\rm star}$}   & 
\colhead{$M_{\rm burst}$} &\colhead{$\tau_{\rm burst}$ }&  \colhead{$M_{\rm sat}/M_{\rm pro}$}&
\colhead{$M_{\rm star}/M_{\rm bar}$}}

\startdata 
S.1&1&9181&1335& 18.75&0.55&4.36&  0.16& 0.12\nl
&&& &33.01&4.06&19.53&  0.23& 0.10\nl
&&& &15.47&1.12&8.61& D&\nl
&2&7310&1059&24.76&0.96&3.53&  1.08& 0.14\nl
& &&&16.52&0.10&1.99& D & \nl
&&& &46.43&4.63&17.60&  0.39& 0.26\nl
&&& &12.38&0.26&5.76& D & \nl
&&& &17.53&1.25&10.11&  0.11& 0.28\nl
&3&6317&917&13.41&0.91&11.50&  0.40& 0.22\nl
&&& &28.89&2.18&11.21& D& \nl
&4&7115&908&29.92&1.44&13.50&  0.16& 0.24\nl
&&& &12.38&0.65&10.39&  0.16& 0.27 \nl
&5&5215&739&8.25&0.49&2.48&   0.43& 0.27\nl
&&& &19.60&0.89&12.4& D&\nl
&&& &22.70&0.91&5.14&  0.22 \nl
&6&6184&798&16.50&1.71&11.50&0.40&0.15\nl
&&&&17.53&1.09&8.11& D&\nl
S.2&1&9199&1307&42.30&0.68&3.94& 0.18 &0.58\nl
&&& &43.33&2.76&13.40& 0.22& 0.61\nl
&2&7383&1071&52.62&1.56&8.86& 1.06& 0.51\nl
&&& &30.96&1.72&6.01& D&\nl
&&& &71.19&4.18&19.08& 0.38& 0.54\nl
&&& &66.93&0.36&8.13& 0.11& 0.59\nl
&3& 6501&918&30.95&1.56&11.21& 0.43& 0.62\nl
&4&7144&921&39.21&1.87&7.47& 0.10& 0.66\nl
&&& &41.12&1.53&13.99& 0.21& 0.46\nl
&5&5294&744&40.24&2.11&15.03& 0.51& 0.78\nl
&&& &33.02&1.43&19.30& D& \nl
S.3&1&3688&550&23.73&0.57&0.81& 0.31& 0.10\nl
&&& &5.16&0.05 &1.23& D&\nl
&2&5665&1056&8.25&0.60&14.25& 0.58& 0.15\nl
&& &&5.16&0.44&13.38& D& \nl
&3&1168&272&22.69&0.43&2.89& 0.35\nl
&&& &6.19&0.18&3.12& 0.10\nl
&4&3302&567&18.57&0.47&2.70& 0.24& 0.26\nl
& && &10.32&0.36&2.76& D& \nl
\enddata

\tablecomments{Units: $[\sigma_{\rm star}] = {\rm M_{\odot} /yr }$;
[$M_{\rm burst}$]=$10^{10} {\rm M_{\odot}}$; [$\tau_{\rm burst}$]=
$10^{8} {\rm yr}$}

\end{deluxetable}

For each stellar burst, we
estimate  the value of a local maximum ($\sigma_{\rm star}$),
its duration ($\tau_{\rm burst}$) and the total amount of stars ($M_{\rm burst}$)
 formed during this period of time \footnote{ Peaks have to be determined
by more than three points higher than $\sigma_{\rm min}$ in order to be
classified as a stellar burst; and their durations, $\tau_{\rm burst}$, are
estimated as the period of time comprised between the first point
to surpass this threshold and the last point which satisfies this
condition. }.
These last two parameters ($M_{\rm burst}$ and $\tau_{\rm burst}$)
 are sensitive to  $f$.
In order to carry out a consistent analysis, once $f$ is chosen, it is kept
constant for all objects in all simulations.
In general, peaks have  $\sigma_{\rm star}> 3\sigma_{\rm min}$.

As already discussed, we have restricted  this study to
 those  peaks that can be directly related to 
merger events,  after the main objects or progenitors are
already formed 
 and better resolved. 
In Table 1 we summarize the principal parameters that characterize those
star formation peaks, including the ratios between the virial 
masses of the progenitor, $M_{\rm pro}$, and 
the satellite
objects, $M_{\rm sat}$, at the time of the merger (i.e., the time when the satellite
enters the virial radius of the parent GLO) and between
the stellar mass content, $M_{\rm star}$, of the system and
its total baryonic mass,  $M_{\rm bar}$,
at the same time.
The ratio $M^{z}_{\rm star}/M^0_{\rm star}$ is the fraction of the total stellar mass
of the progenitor GLO at $z=0$ that has actually been formed at the red-shift of each merger
 ($M^{z}_{\rm star}$).
 GLO 6 in S.2  has not been included since it is not
possible to clearly isolate the star formation peaks associated with the
mergers due to the high level of noise in its SF history. 
During some merger events, two stellar peaks have been detected.
They are denoted by a letter D in Table 1 and will be discussed in more detail
in a separate paper (Tissera et al. 2000).
Note that we study a total number of 25 different merger events. So,
even if the GLOs whose evolutionary 
histories are analyzed are restricted to the more massive ones, we look
at all mergers recorded in their merger trees. Hence, this is
equivalent to have  performed 25 different mergers using
controlled toy-models, with the advantage that each one of
these mergers has  physical properties determined by the underlying cosmology
and the astrophysical model, and occurs at different stages of evolution.
Unfortunately, this sample is not large enough to study
the possible dependence of the stellar burst characteristics
with the red-shift.

Let us now try to investigate if the parameters that characterize the
stellar bursts are consistent with observations of galaxies undergoing
strong stellar activity at different red-shifts
and how they change with the model parameters ($\sigma_{8}$,
$c$, $T_*$).
As can be seen from Table 1, the values of $\tau_{\rm burst}$, $M_{\rm burst}$
and $\sigma_{\rm star}$ vary among simulations and
are different for  S.1 and S.2 versions of the same merger
event. 
In Table 2 are shown the mean values $<\sigma_{\rm star}>$, $<\tau_{\rm burst}>$, and $< M_{\rm burst}>$
in units of ${\rm M_{\odot}/yr}$, $10^{8} {\rm yr}$ and ${\rm 10^{10} M_{\odot}}$,
respectively.
As expected, the higher values are measured for peaks in S.2 
since, as 
discussed in Section 2, the gas is transformed into stars
more efficiently in this simulations than in S.1 and S.3. 
Recall that the only difference between S.1 and S.2 is the
$c$ parameter, and, that S.1 and S.3 have different 
bias parameter and critical temperature, $T_*$,  but equal  $c$ value. 
In S.3, at lower $z$, the  stellar peaks are less
important when compared to those in S.1, 
 because a larger fraction of gas has been consumed at higher $z$ (so
the gas density is lower at later times) and also because of
the more restrictive
temperature criterium $T_*$ used.
 At those low $z$, GLOs in S.1 
are more gas-rich and can produce stronger stellar bursts.
For simulation S.3, we see that the mean stellar mass, $<M_{\rm burst}>
= 3.90 \times 10^{9} {\rm M_{\odot}}$, and the starburst time-scale,
$<\tau_{\rm burst}> =  5.14 \times 10^8 {\rm yr}$, are consistent
with observed values inferred for starburst galaxies and high
$z$ objects undergoing important star formation activity
(see Sawicki \& Yee 1997; Kennicutt 1998 and references therein). Also note that these values depend on $f$.  Had we 
chosen a higher threshold ($\sigma_{\rm min}$), $\tau_{\rm burst}$
and  $M_{\rm burst}$ would have been smaller in all cases. 
Note also that the simulated bursts occur at different $z$, so
a direct comparison with observed objects at high $z$ is
complex since, for example, the effects of dust reddening are still
uncertain and lead to underestimations of the SFRs. And, on 
the other hand, a comparison with only local starburst galaxies could
be misleading, since the properties of the galactic objects 
change with $z$ and so could do their SF histories (e.g., \cite{guz97}; \cite{dri98}; \cite{flo98}).  

\begin{deluxetable}{crrr}
\footnotesize
\tablecaption{Mean values for the parameters of the stellar bursts. \label{tbl-1}}
\tablewidth{0pt}
\tablehead{
\colhead{S} & \colhead{$<\sigma_{\rm star}>$}   & \colhead{$<\tau_{\rm burst}>$}
   &
\colhead{$< M_{\rm burst}>$}
}

\startdata
1& 20.79  &9.28  &1.36\nl
2& 40.02  &11.49  &1.79\nl
3& 12.51  &5.14  &0.39\nl

\enddata
\end{deluxetable}



The next question is in relation to the  factors  triggering
 these stellar bursts and
if cause-effect relations can be isolated.
We  estimate the ratio between the virial 
mass of the satellite ($M_{\rm sat}$) that is falling in, and
the virial  mass of the progenitor ($M_{\rm pro}$) just
 before the satellite enters
the virial radius of the parent galaxy. 
We plot  $\sigma_{\rm star}$ vs. $M_{\rm sat}/M_{\rm pro}$, including the double peaks
(see Section 3.3)
in Figure 3a.
The response is different depending on the simulation.
In simulation S.2, $\sigma_{\rm star}$ values are higher
than those in S.1  although 
the GLOs are   numerically identical have the same merger trees
 in both simulations.
This difference is originated in the different star formation
efficiency as already discussed: it is easier to transform a gas particle into 
a  star one in S.2 than in S.1. The
same argument is valid for S.3  which has
the lower values for roughly 
the same $M_{\rm sat}/M_{\rm pro}$ ratios. Note that events
with approximately equal  $M_{\rm sat}/M_{\rm pro}$ ratios produce different $\sigma_{\rm star}$
in the same simulation.

An interesting  fact observed from
 this figure is that, even a merger
with a low mass satellite, $M_{\rm sat}/M_{\rm pro}\approx 0.10$, 
correlates with
an increase of the star formation rate in all simulations,
 with $\sigma_{\rm star}$
values similar to those corresponding to higher $M_{\rm sat}/M_{\rm pro}$ .
This result is in agreement with
\cite{mh96} who studied in detail the accretion of low
mass satellites and found that even  a merger with an object with a mass
 of 10\% the parent galaxy mass,
can produce an inflow of gas to the center of the main object  
fueling  a starburst. This result may suggest that a comparable 
companion is not a  necessary condition for triggering star formation, but
a smaller one may produce significant effects.

The availability of gas in condition of forming stars is one necessary
condition to trigger a stellar burst. Hence, it would be expected 
a correlation  between the gas content of the system and the strength
of the stellar peaks.
In Figure 3b, we plot  $\sigma_{\rm star}$ vs. $M_{\rm gas}/M_{\rm bar}$
where $M_{\rm gas}$  is the total gas mass of 
the satellite and its progenitor within $r_{200}$,
 before the merger,
and their total baryonic  mass $M_{\rm bar}$
 (gaseous and stellar 
masses of the satellite and the progenitor together at $r_{200}$)
 at the same time,
for the objects in S.1,
S.2  and S.3. 
It can be seen that there is no correlation.
Note also that not always the  more gas-rich objects have the larger
$\sigma_{\rm star}$ or that equally gas-rich objects have 
different $\sigma_{\rm star}$, even  in the same simulations, that is,
with the same SF parameters. 
However, the absolute value of the bursts depends on $c$:
S.2 has higher $\sigma_{\rm star}$ values even though the objects are more
gas-poor than their counterpart in S.1. 
Note also that within the same simulation, GLOs have very similar
gas abundances. Hence the fact that equal massive mergers produce different
$\sigma_{\rm star}$ in the same simulation cannot be directly related
to a difference in gas richness of the objects involved.

Given the  durations  of the bursts, $\tau_{\rm burst}$,
and the total stellar
mass formed during that period, $M_{\rm burst}$,
it is possible to estimate an overall
 star formation rate $<SFR>_{\rm burst}=M_{\rm burst}/\tau_{\rm burst}$
associated to the
burst. We found values ranging from
 $50 \ \rm{M_{\odot}/yr}$
to $2 \ \rm{M_{\odot}/yr}$ depending on the simulation.
For illustration purposes, we plot them against the the ratio
$M_{\rm sat}/M_{\rm pro}$ (Figure 3c). Again, we found
no correlation signal.
 Since $<SFR>_{\rm burst}$
depends not only on the amount of stellar mass formed, but also on
 the duration
of the burst, this lack of correlation is not surprising.
The duration of the bursts, $\tau_{\rm burst}$,
 can be determined by different parameters
such as orbital orientation, internal structure, star 
formation efficiency, in a 
complex way.
From this figure we can also observe that a merger with a satellite of
$0.10$ or $0.40$ the mass of the parent object may produce the same
average $<SFR>_{\rm burst}$ within the same simulation.
This fact supports the idea that
 it is not only the masses of each component in each halo that matters,
but other factors could also be relevant, such as  the dynamical characteristics of the encounter and the structural
properties of the baryonic clumps that merge.
We  have also searched for possible correlations between both, $\tau_{\rm burst}$
and $M_{\rm burst}$, with
 $M_{\rm sat}/M_{\rm pro}$ and $M_{\rm gas}/M_{\rm bar}$.
No signal was detected implying that the total stellar mass and
the burst duration are not simple functions of only the relative masses
of the merging objects or their gas richness.
However, they  do depend on $c$ 
as already pointed out (see Table 1).

These results prompt us to differentiate between minor and major mergers
in order to ascertain if a  hidden effect could be disentangled.
In Figure 4a we plot again  $\sigma_{\rm star}$ vs. $M_{\rm gas}/M_{\rm bar}$ but,
in this case, filled symbols represent major mergers, while open ones,
minors events (circles, triangles and
pentagon for S.1, S.2 and S.3, respectively). The limit between major and minor mergers has been 
set at $M_{\rm sat}/M_{\rm pro}=0.35$ (\cite{bau96}).
We include single bursts and the first components
 of double ones
(secondary components are
formed with the  remanent gas  after the first one, so actually they
depend on the properties of the first component).
We can see from this figure that a minor merger can trigger
a burst as strong as a major one, even if the systems is
less gas-rich, and that a major merger, in some cases, does not
trigger a strong burst even in gas-rich collisions.

To deepen into the burst process, we compare the amount of stars
formed in a given stellar burst,  $M_{\rm burst}$, with
the amount of gas available in the system to form stars, $M_{\rm gas}$.
In Figure 4b we do not see any clear correlation, that is,
the amount of gas available at the beginning of the merger that is actually
transformed into stars does not mainly depend on   $M_{\rm gas}$.
In most cases, it is smaller than the amount of gas available
in the system, so  the gas mass is not actually completely
exhausted during a burst.  This can also be  seen in the SFR
histories, from where we observe that the SF continues after a
burst, although at a lower rate.


An estimate of the efficiency
of the star formation process in each burst can be defined as the fraction
of available gas in the system, at the time
the satellite enters the virial radius of the progenitor, that was actually converted into stars, $M_{\rm burst}/ M_{\rm gas}$.
We plot this burst efficiency ratio versus $M_{\rm gas}/M_{\rm bar}$
(Figure 5a) and 
$M_{\rm sat}/M_{\rm pro}$ (Figure 5b) for the single bursts and
for the first peaks of double bursts.
As can be seen from these figures, there is no
correlation between the burst efficiency  and
the gas abundance of the GLOs,
implying that, in these GLOs, the burst efficiency is
not determined by the gas abundance (neither is there   a correlation
of the efficiencies
with $M_{\rm gas}$, which suggest that they are doubtfully
 determined by numerical resolution). 
But a trend is present with $M_{\rm sat}/M_{\rm pro}$,
suggesting that massive mergers can induce  more efficient transformations
of the gas into stars. 
The mean values of $M_{\rm burst}/ M_{\rm gas}$ for stellar bursts
associated with minor ($M_{\rm sat}/M_{\rm pro} < 0.35$) and 
major ($M_{\rm sat}/M_{\rm pro} \geq 0.35$) mergers  for S.1, S.2
and S.3 are  (0.24, 0.35),
(0.23, 0.72) and (0.10, 0.19), respectively. 
We have averaged all peaks regardless of the red-shift at they have occurred.


This trend has been reported by Mihos \& Hernquist (1994, 1996)
 who used high-resolution models
of pairs of merging galaxies, and successfully implemented by
Somerville, Primack \& Faber (1998) in a semi-analytic model.
We found the same relation for objects formed in a cosmological
context and for a  range of merger parameters that naturally arises 
in coherence with the stage of evolution of the GLOs.
However, our efficiencies are not as high as those  claims by
Somerville et al. (1998) and depend on the SF parameters.

All these results suggest that, at least, a third parameter is
playing a role in the triggering of the bursts: it is not
enough to have available gas in the system, independently of
the relative mass of the colliding objects. The gas has
to be violently compressed in short time-scales in
order to induce a starburst, and in this case, mergers seem
to be doing part of the work.

Concerning numerical resolution, the lack of correlation found between
the burst characteristics and the gas abundance ($M_{\rm gas}/M_{\rm bar}$)
 or the gas mass ($M_{\rm gas}$, that  gives a rough idea of the gas
resolution of the system at the time of
the merger) strongly suggest that they 
are not determined by numerical
resolution.
 A second fact that supports this point is that
 peaks in S.2 have
higher $\sigma_{\rm star}$ values than those of the same GLOs in S.1, 
even though the objects are more gas-poor than those in S.1.

Because the SF process depends directly on the gas density, its correct numerical
description is crucial. As shown by Tissera \& Dom\ci nguez-Tenreiro (1998)
and Dom\ci nguez-Tenreiro et al (1998), the gas density within
the dark matter halos of  massive systems in these simulations
 are rather well described.
In this sense, an advantage of these simulations is that the mass
of gas and dark particles are equal, implying that the dark matter
is resolved with a factor of about 10 more particles than the gas.
This fact assures that two-body effects are unimportant
and that the dark matter profiles are
 well represented in the central regions.
An adequate resolution of the dark matter profiles strongly helps the gas
to cool and collapse inside the central regions following a correct
density profile (\cite{sw97}). This implies that
the SF process, that depends on the gas density, can be also
adequately followed. 


\section{Stellar Population and Color Distributions}

As already discussed in Section 3, the star formation history of each
individual galactic object can be followed with look-back time.
This information can be combined with  stellar population synthesis models
 to estimate the luminosities
and colors of galactic objects throughout their
evolutionary history (\cite{tis97}).

We use the models of   Bruzual \& Charlot (1993) to
calculate the luminosity of a particle as a function of wavelength
 $\lambda$ and
$z$. We assume a
Miller-Scalo initial mass function with a lower mass cutoff of $0.1 M_{\odot}$
and an upper mass cutoff of $125 M_{\odot}$, and  a burst 
of 20 time-steps duration ($2.8 \times  10^{8}$ yr for S.1 and S.2, and 
 $2.4 \times 10^{8}$  for S.3)  for
each particle.
Then, we sum up the luminosities of the particles belonging to an object
and, from the total luminosities, we estimate their colors and magnitudes at
 different wavelengths.
No supernova energy
injection or metallicity
enrichment have been included.
We have neither included reddening effects so some
comparison with observed data may result to be rather crude and
unfair to simulated colors.

In Figure 6 we plot $U-B$ vs $B-V$ at $z=0$ for the simulations
analyzed: S.1, S.2  and S.3.
We also include the observational data obtained from
the RC3 catalogue (\cite{deV91}) which has UBV photometry,
Hubble types and measured red-shifts for $\sim 10000$ galaxies. We plot  $U-B, B-V$
locus for both early  and late-type
 galaxies with $z\le 0.05$ in RC3.
We see that most of the simulated GLOs are bluer than
the galaxies although the maximum departure from the observations
is at most of $ \approx 0.3$ magnitudes. This is clearly due to fact that the
star formation rate histories of the simulations
 produce higher rates than observed at $z< 0.05$.
But in spite of this fact, a better agreement between simulated
and observed colors is not impossible and it only requires 
a lower star formation rate at $z\approx 0$ which can be accomplished, for example,
by using a lower $c$ value or including  SN feedback.
It is also possible that, because at high $z $, GLOs are not very well
resolved in these simulations, the gas is inefficiently consumed into
stars leading to more gas-rich GLOs at lower $z$. On the other
hand, higher resolution
experiments without a correct  feedback implementation could lead
to a very effective SF process at high $z$ producing GLOs with lower
SFR and redder colors at $z=0$ (e.g., Steinmetz \& Navarro 1999).
In our simulations, low resolution at very high $z$ acts as a feedback effect.

Note that the SFR histories among the simulations are quite different
(because of the different SF parameters used), however
the distributions of colors at $z=0$ on this color-color diagram
are very similar. Hence these colors are very insensitive to
the star formation history of the GLOs and are not a very safe way of
adjusting SF parameters in semi-analytic or numerical models.

In Figure 7 we plot $B-V$ vs $z$   for
four galactic-like objects in simulations S.1  and S.2.
Although the number of outputs available is small, it is
still possible to follow the color evolution of the progenitor.
It can be clearly seen  that the color path is not smooth.
An object can move from blue to red  and
 back to blue colors depending on their history.
The change in colors in these models can be as
large as half a magnitude.
In particular,
each change in direction in the color-color
diagrams correlates with a peak of star formation as  expected.
The strength in the change depends on the relative number of the new
to the old stars.
In this figure we have plotted the tracks of objects in S.1 and S.2 with
the aim at comparing the influence of the SF efficiency. As can be
seen, although the general behavior is similar, the detailed
evolution is different.
The $B-V$ tracks of GLOs in S.2
are  displaced in time with respect to those in S.1. Since the only difference
between these simulations is their star formation efficiency parameters, 
the  color evolution
of the objects could be affected by its choice.
 
An even clearer approach for analyzing the evolution of the stellar population
and its relation with mergers is to look at $B-I$ vs $z$. 
In Figure 8 we plot it for the same four objects in S.1 shown in Figure 7 (solid lines for a Miller-Scalo IMF).
This figure
shows clearly when there is a burst of star formation and its correlation with
mergers (the red-shift at which the satellite enters
the virial radius of the progenitor has been indicated with
an arrow pointing up, while  the actual
fusion of the baryonic cores has been indicated with
an arrow pointing down).
In some cases, because of the small number of outputs, we could have
missed the peak and just see the remnants. Note that this problem is not
present in the star formation history since it is saved completely 
until $z=0$. 
Again, the changes in color depend on the particular characteristic of
the mergers and the proportion of old to new stellar populations.
We also estimate colors using a Salpeter IMF with the same
lower and upper cutoffs (dashed lines).
As can be seen the differences are very small, thou the colors calculated
using  Salpeter IMF tend to be redder.
Unfortunately, the smaller stellar mass   allowed by Bruzual \& Charlot's models
is $0.1 M_{\odot}$, so we could not evaluate the effects of 
assuming a lower mass cutoff such as $0.01 M_{\odot}$.
In Figure 9, we plot $B-I$ vs $z$ for the  4 GLOs shown in Figure 7 and 8
(simulations S.1 and S.2)
and observed values from the LDSS2  
(\cite{ell96})
and LRIS (\cite{guz97}).
Despite the $\rho_{\rm SFR}$ are very different between these simulations,
their evolutionary color tracks are in general agreement  with observations. 

Obviously not all star bursts are triggered by mergers, but, since
they are common events in a hierarchical scenario, their possible influence
on the evolution of colors cannot be ignored.
This picture deduced from the star formation histories and 
 color evolutionary tracks
 resembles the 'Christmas Tree' model discussed
by Lowenthal et al. (1997) in which individual star forming blobs come
and go. According to our models,
 the star formation in this blobs would be the result of two
contributions: one due to an approximately constant ambiance star formation and
a certain number of SF peaks.
The aggregation of substructure
according to a hierarchical scheme would be one possible mechanism of 
SF triggering.
The parent galaxy will evolve by undergoing a number of mergers which may
trigger starbursts depending on the particular characteristic of the
encounters, and  the physical properties of
the objects involved.
Hence during violent phases colors would became bluer to 
change again to redder ones as a quiescent  period takes place.
Moreover, because of the way the structure forms in a hierarchical
scenario,
the higher
the $z$, the higher the probability that  massive   objects 
 would be 
observed to be undergoing  an important star formation
activity period (\cite{guz97}) since the rate of mergers
increase with $z$, GLOs are gas richer and the gas is denser.

This technique that combines hydro-dynamical simulations and synthesis 
evolutionary models has proved to be potentially powerful to study
the formation of galaxies as a function of $z$. Nevertheless, more
complex models will be needed in order to mitigate numerical
resolution problems, and to allow us to numerically follow the 
evolution of galactic structure of different  masses with look-back time.

\section{Summary}

We have analyzed the history of star formation in galactic objects simulated
within the framework of a clustering hierarchical model.
Our aim was to use a set of three cosmological simulations  to 
study the possible interplay between hierarchical 
aggregation and star formation.
We found that, if the structure in the Universe is well represented by 
a hierarchical clustering model, then our results suggest that the
process of aggregation of substructure could be one of the mechanisms 
that triggers star formation in galactic systems.
In this work, SN effects have not been included. It is expected that
they will contribute to set a self-regulated SF. However, their
actual impact on galaxy scales remains to be clearly stablished.

Our conclusions can be summarized as follows:

I. The star formation rates as a function of $z$ of our simulated GLOs have
two components: one approximately constant (${\rm ASFR} < 3 $ ${\rm M_{\odot}/yr}$), and
a series of stellar bursts superposed.
We found that the
aggregation of  substructures by the progenitor objects correlates with
the presence of stellar bursts. These bursts
 last $ 10^{8}-10^{9} {\rm yr}$ and produce stellar masses  
of $ 10^{9}-10^{10}{\rm M_{\odot}}$. For S.3, these
parameters are   consistent with observations of 
 galaxies undergoing strong SF activity at different $z$.
For S.1 and S.2 the values are higher as could be predicted
from the global $\rho_{\rm SFR}$.

II. No correlation between  the strength of the stellar peaks, 
its duration  and  stellar mass formed, on one hand,  and the 
ratios $M_{\rm sat}/M_{\rm pro}$ and $M_{\rm gas}/M_{\rm bar}$,
on the other hand, was found. This  fact
implies that they do not determine the characteristics of the burst by their own. 
And that the strength of a stellar peak can not be predicted 
only from the gas richness or the size of the colliding satellite.
Mergers with equally massive objects produce different effects in the same
simulation.
When major and minor mergers are distinguished it can be seen  
that both  
of them can  generate different stellar maxima
 regardless of their mass
content. The strength of the star formation peaks, however, do depend on the SF
efficiency parameter used in the models.

III. We found a trend for massive mergers
to be more efficient at inducing a transformation of the
gas available in the system into stars.

IV. In agreement with Hernquist (1989b), we find that 
a merger with a satellite of even $10\%$ of the progenitor
mass can be correlated with a stellar burst independently
of the value of SF efficiency used. 
This result would imply that, when searching for  a companion as
the triggering factor of strong star formation activity in a galaxy, not only
similar galaxies should be checked out but, also smaller ones
(\cite{dp97}).

V.
The color tracks of  GLOs are  not smooth, but go from 
bluer to redder in the quiescent phases of evolution,
and vice-versa in the violent phases corresponding to mergers.
The amount by which colors change on the latter depend
mainly on the star formation 
history of each GLO. But color distributions at $z=0$ are
quite insensible to their particular star formation histories.

\acknowledgments

The author is  grateful to Prof. Diego G. Lambas, Rachel 
Somerville and the anonymous referee of this paper  for stimulating discussions and comments.
We thank Francois Hammer for providing useful information. 
P.B.Tissera thanks Imperial College  the University of Oxford and the Centro de Computaci\'on Cient\ci fica
(Universidad Aut\'onoma de Madrid) and the University of Oxford
for providing the computational support for this work and for their hospitality.
This work was partially supported by DGES (Spain), through grant
PB96-0029 and Consejo Nacional de Ciencia y Tecnologia (Conicet, Argentina). 


\clearpage

\figcaption[]{ Cosmic star formation rate density, $\rho_{\rm SFR}$, versus
$z$ for simulations S.1 ({\it dotted lines}), S.2 ({\it dashed lines})
and S.3 ({\it solid lines}). We also include observational
values reported by Gallego \eti (1995, {\it asterisk}), Madau \eti 1996 ({\it filled
triangles}), Lilly \eti (1996, {\it open
squares}), Connolly \eti (1997, {\it ten arm stars}),
Sawicki \eti (1997, {\it open stars}), Hughes \eti (1998, {\it filled square}),
Flores \eti (1998, {\it crossed open squares}) and Steidel \eti (1998,
{\it crossed open cricles}). 
\label{f}}

\figcaption[]{ Star formation rate history of four
 galactic-like objects analyzed in
simulation S.1 as a function of look-back time. The horizontal solid line represents
the ambient star formation rate $\sigma_{\rm min}$ for each object.
The time at which the satellite enters
the virial radius of the progenitor has been indicated with
an arrow pointing up, while  the actual
fusion of the baryonic cores has been indicated with
an arrow pointing down.
Globally, the SFR of the GLOs can be described by
the superposition of a constant star formation rate
and a series of bursts.}


\figcaption[]{ The local  maximun of star formation $\sigma_{\rm star}$ versus
the ratio between the virial masses of the satellite and
the progenitor, $M_{\rm sat}/M_{\rm pro}$ (a) and
the ratio between the gas content and the total baryonic mass of the
colliding system, $M_{\rm gas}/M_{\rm bar}$ (b), and the average star
formation rate $<SFR>_{\rm burst}$ versus $M_{\rm sat}/M_{\rm pro}$ (c),
  for starbursts identified in galactic-like objects
in simulations S.1 ({\it open circles}), S.2 ({\it filled triangles}),
and S.3 ({\it filled pentagons}; for the sake of clarity,
 we have not included in figure c
the data corresponding to the first burst of GLO 1 in S.3.:
 $M_{\rm sat}/M_{\rm pro} = 0.31$ and  $<SFR>_{\rm burst} = 70.37 
{\ \rm M_{\odot}/yr}$.}

\figcaption[]{ a) $\sigma_{\rm star}$ versus   $M_{\rm gas}/M_{\rm bar}$
 for GLOs in
S.1 ({\it circles}), S.2 ({\it triangles}) and S.3 ({\it pentagons}). Solid symbols
represent major mergers while open ones, minors. b) Stellar mass
formed during a burst, $M_{\rm burst}$, as a function of the total
gas mass available in the system, $M_{\rm gas}$ (symbol code as
in Figure 3). Single bursts and first components of double ones
have been included.}

\figcaption[]{ The fraction of gas converted into stars in a star formation burst,
$M_{\rm burst}/M_{\rm gas}$, versus the gas fraction of
the system $M_{\rm gas}/M_{\rm bar}$ (a) and the relative masses of the colliding
objects  $M_{\rm sat}/M_{\rm pro}$ (b), for single peaks and the first
components of double ones.
A trend is found for  massive mergers to produce more
efficient bursts.}



\figcaption[]{ $U-B$ vs $B-V$ diagram for GLOs at $z=0$ in S.1 ({\it open circles}),
S.2 ({\it filled triangles}) and S.3 ({\it filled pentagons}).
We also plotted colors for early ({\it five arm stars}) and late ({\it ten arm stars})
types galaxies from  the RC3
 catalogue.}

\figcaption[]{$B-V$ colors versus $z$ for four objects in
simulations S.1 ({\it open circles}) and S.2 ({\it filled triangles}).}

\figcaption[]{$B-I$ colors versus $z$ for the same four objects in simulations
S.1 estimated by using two different IMF: Miller-Scalo ({\it solid line}) and
Salpeter (x=1.5; {\it dotted line}).}

\figcaption[]{Comparison between the $B-I$ tracks of
GLOs in S.1 ({\it dashed lines})  and S.2 ({\it solid lines})
 with observation data: LRIS ({\it filled circles})
and LDSS2 ({\it open circles}).
}
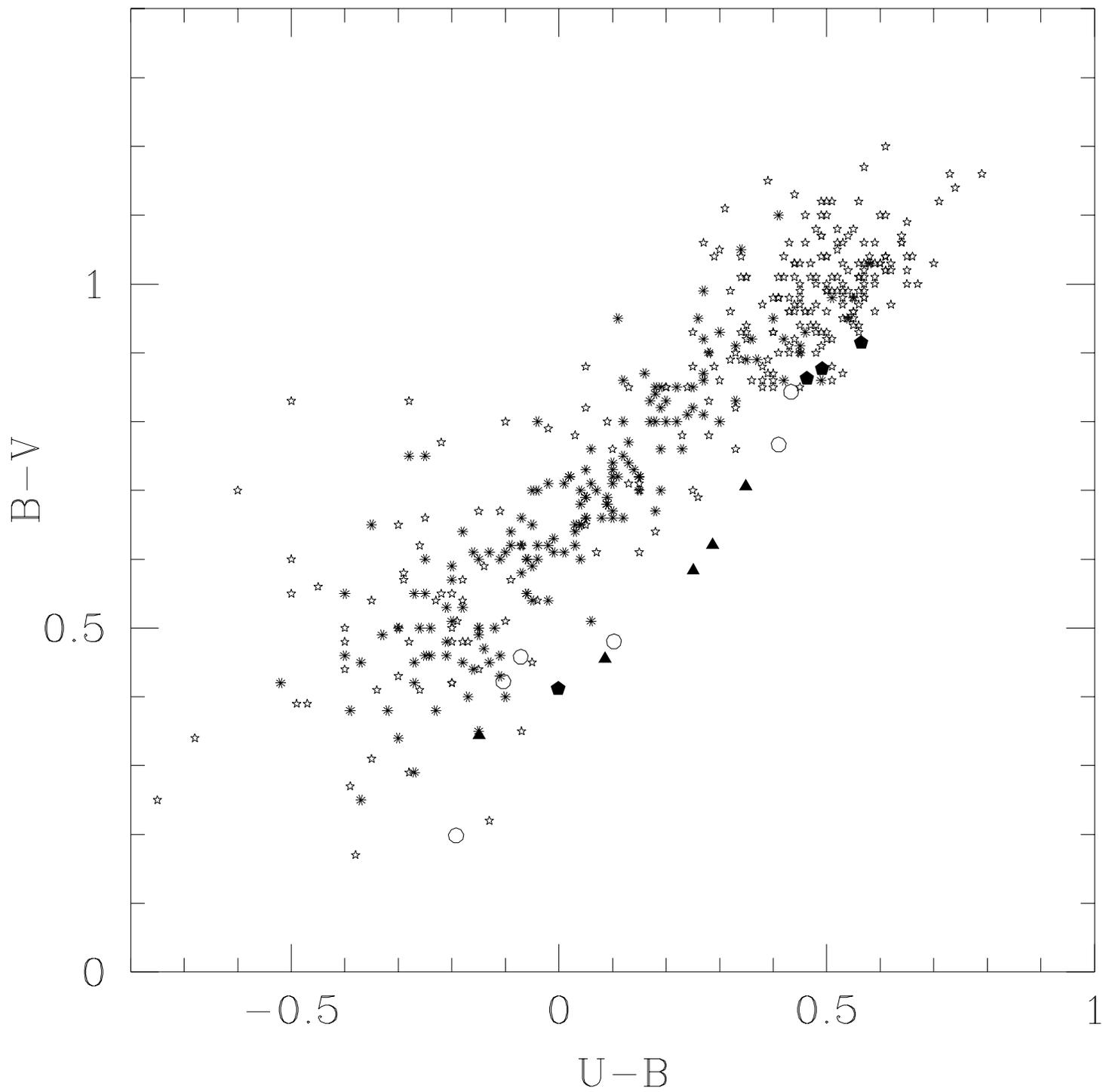
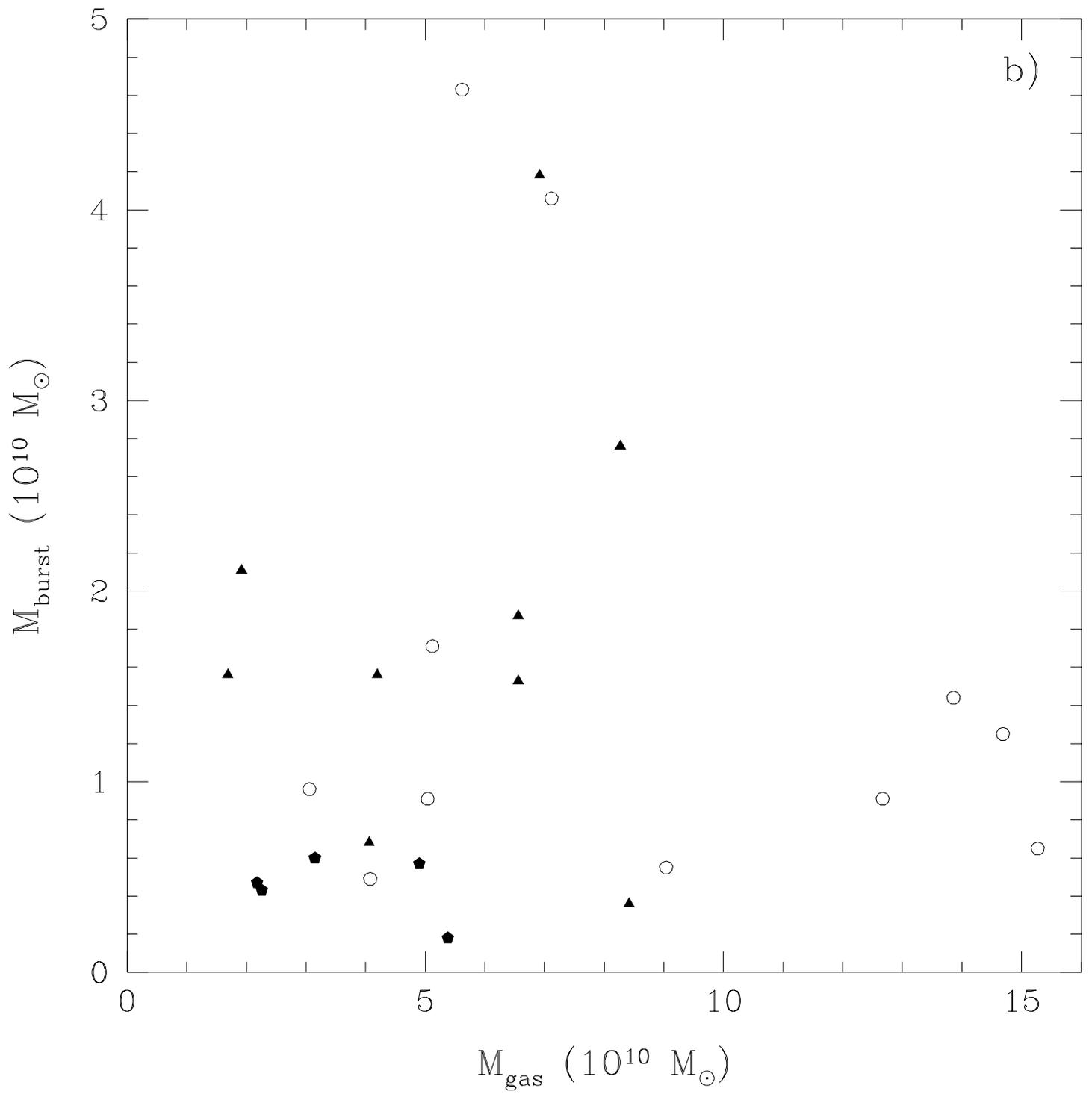
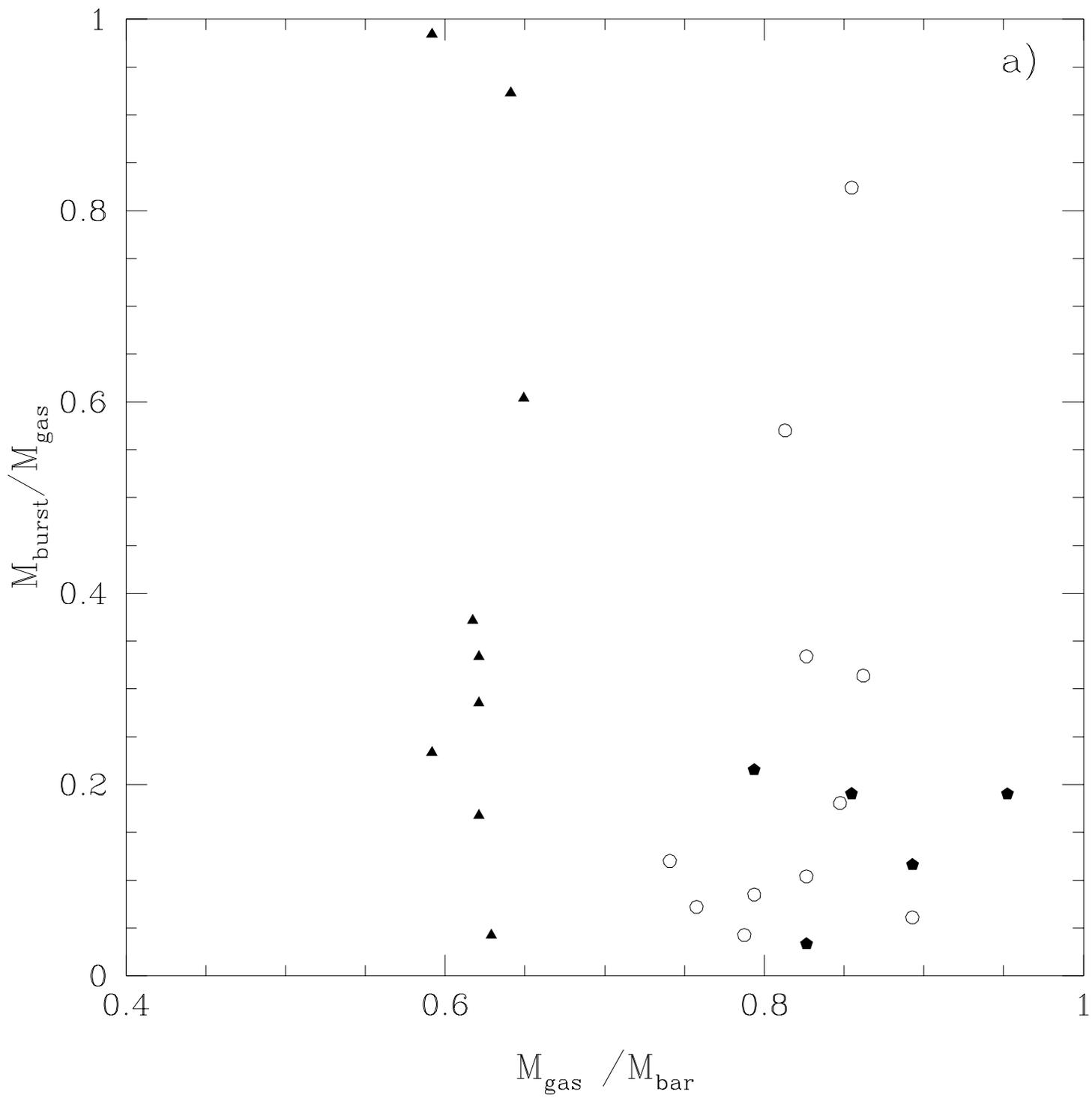
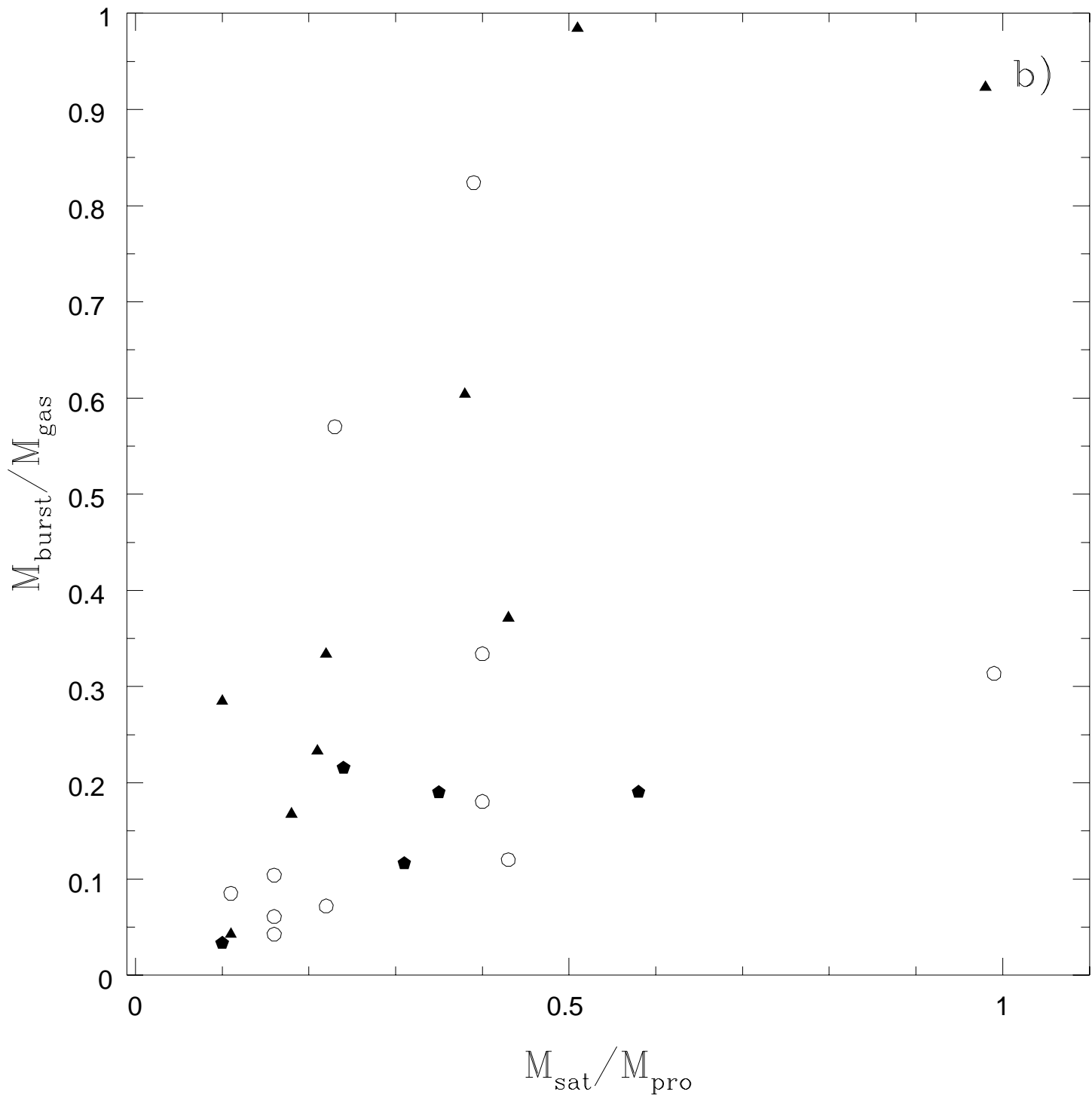
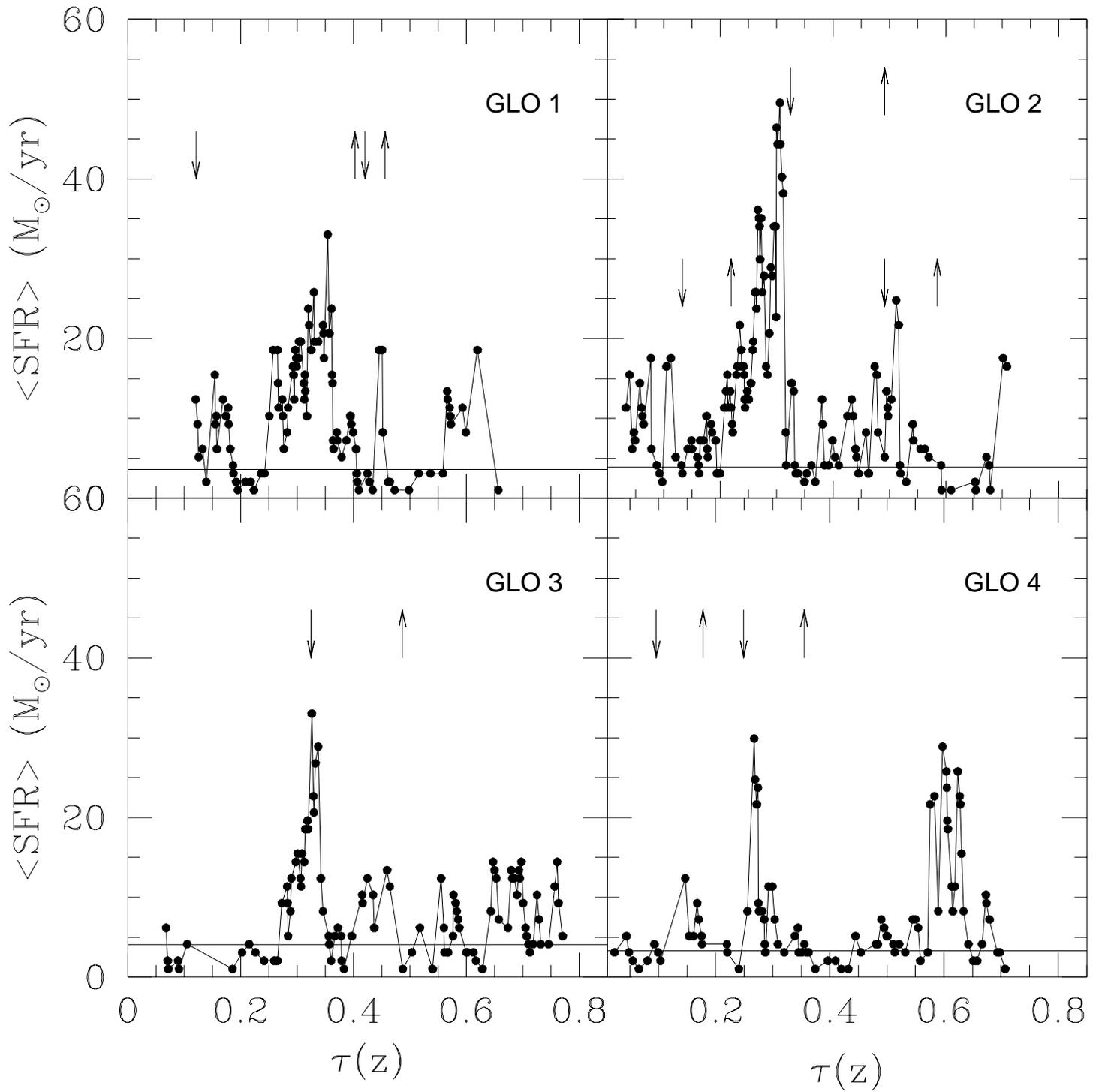
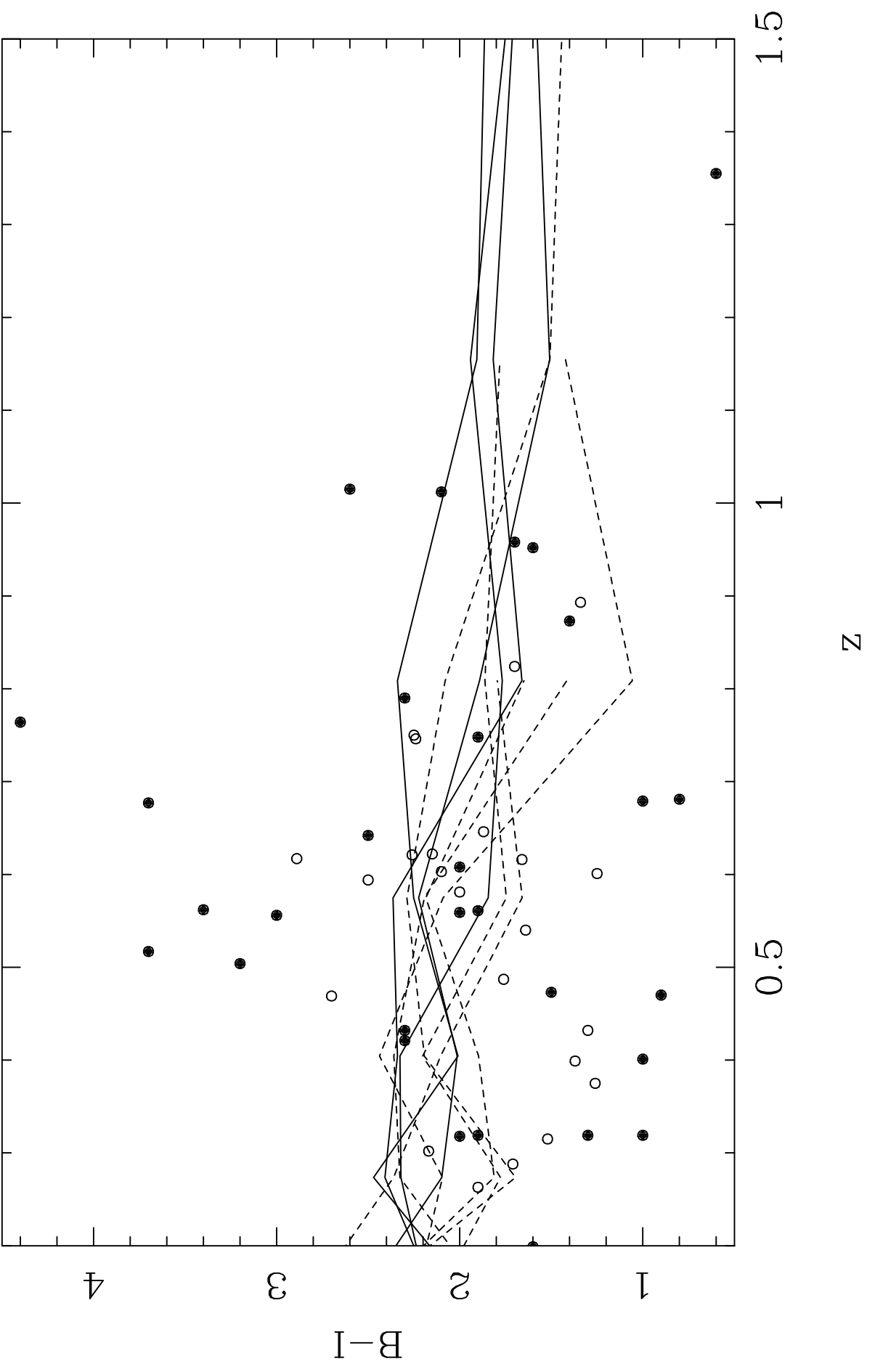
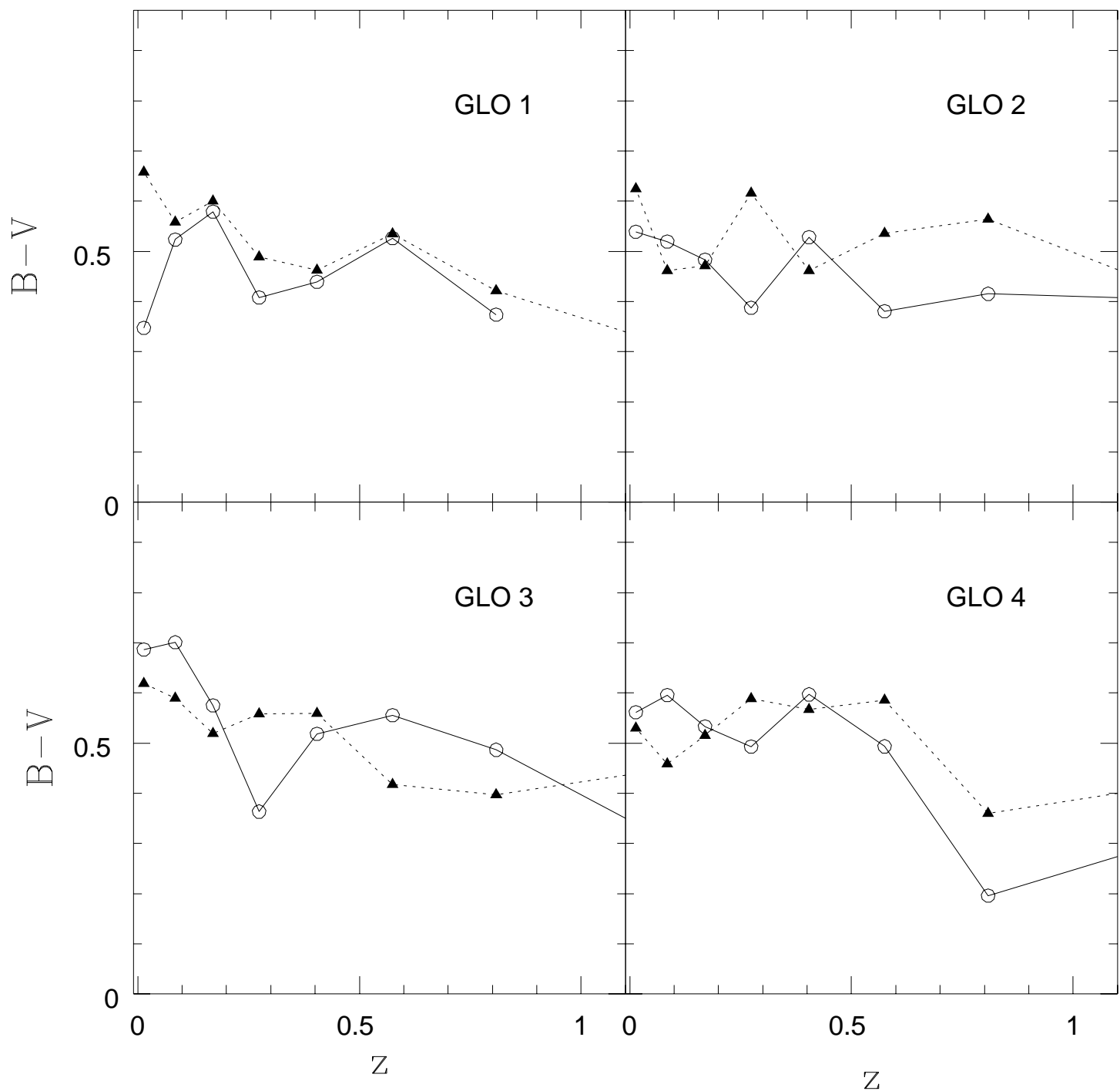
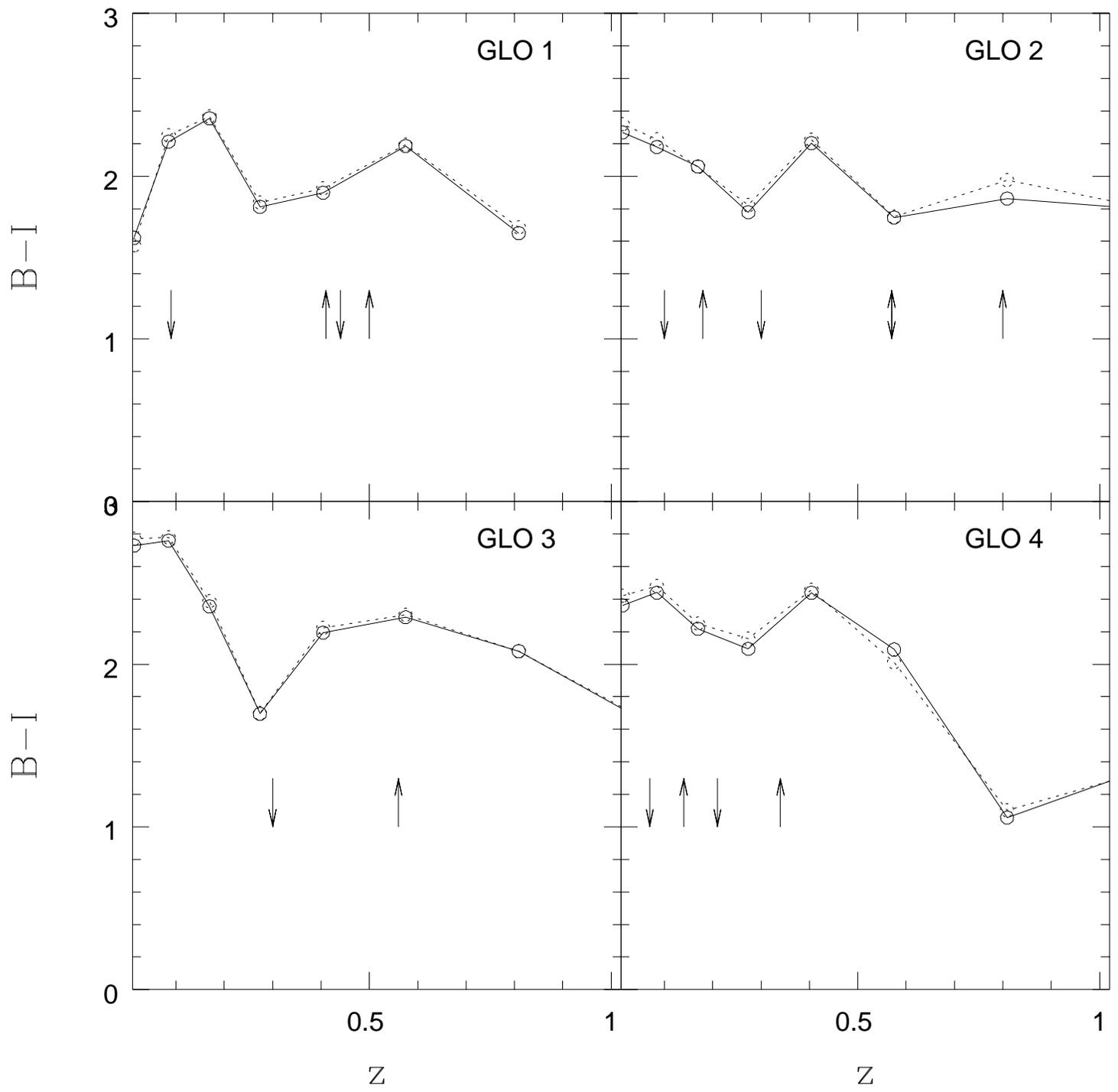
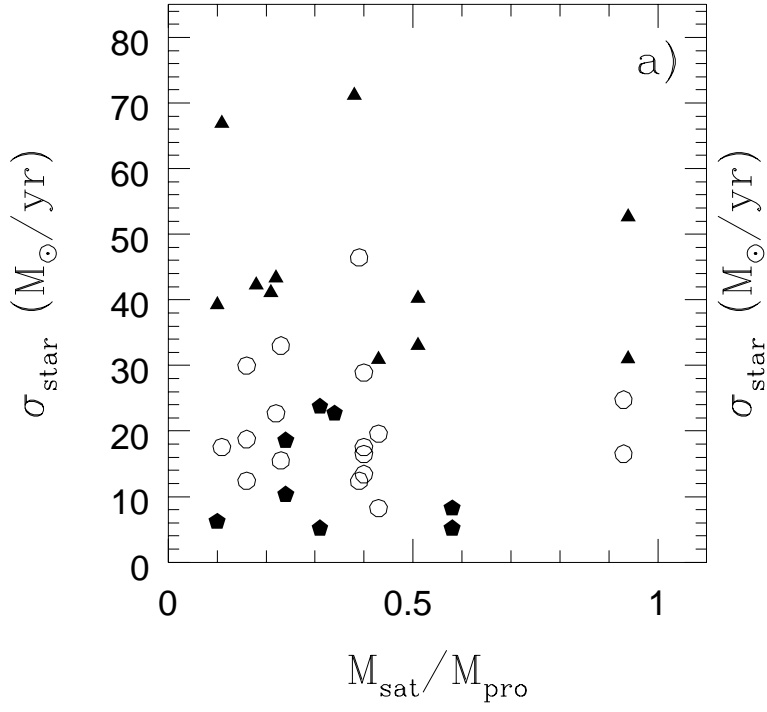
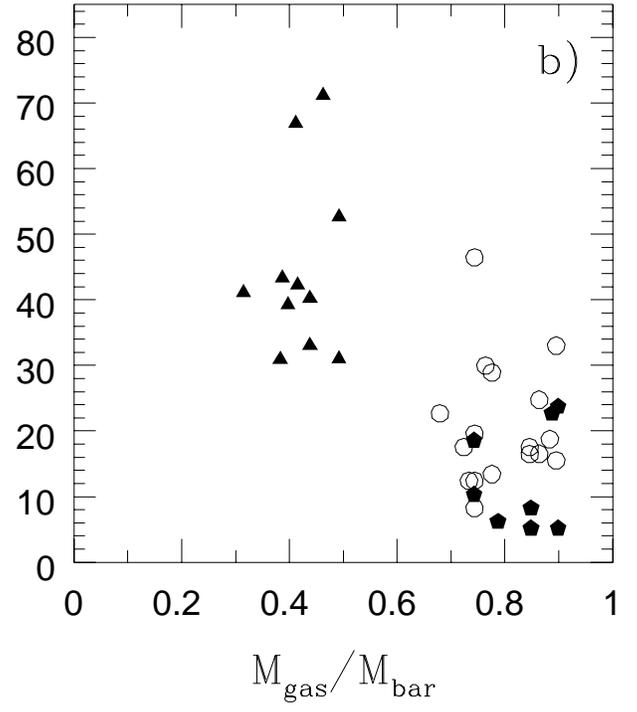
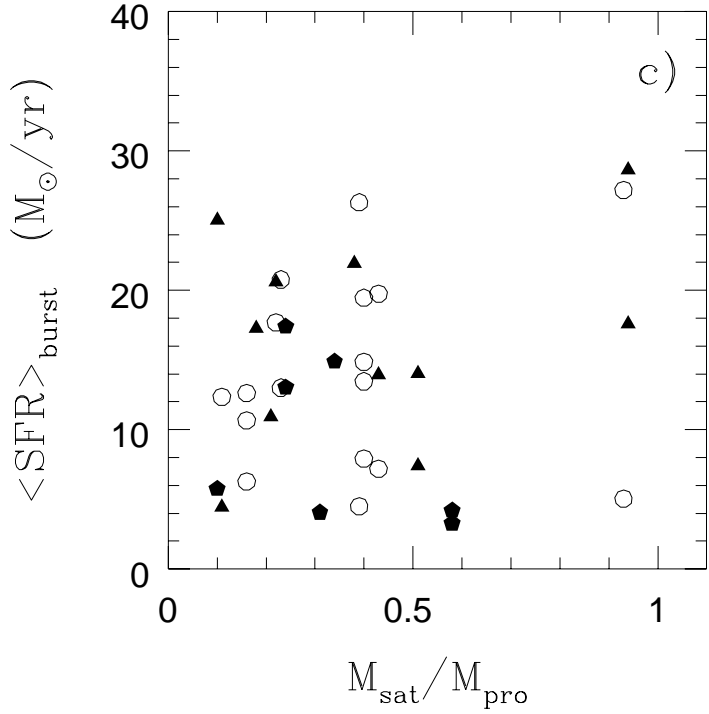
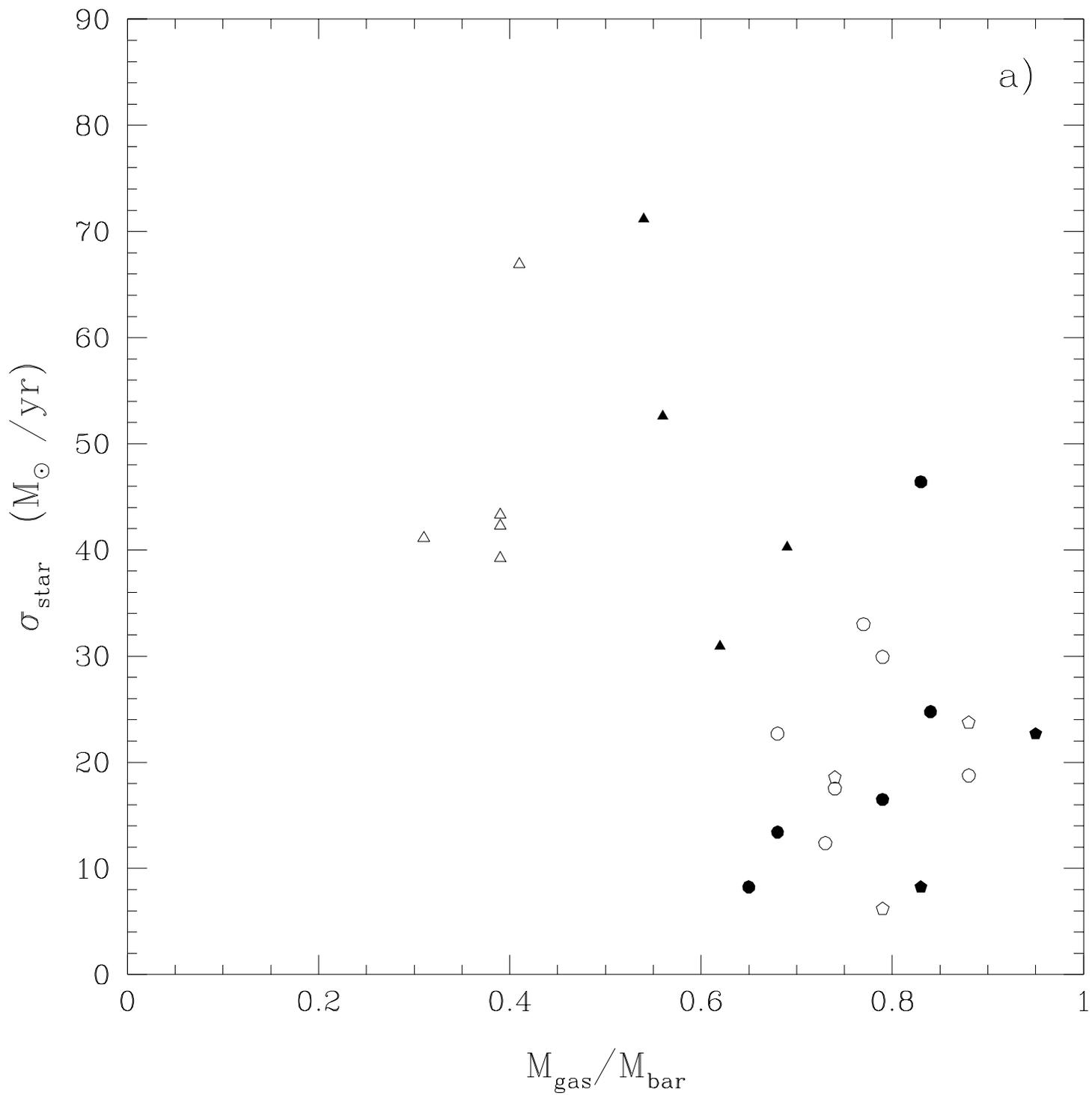
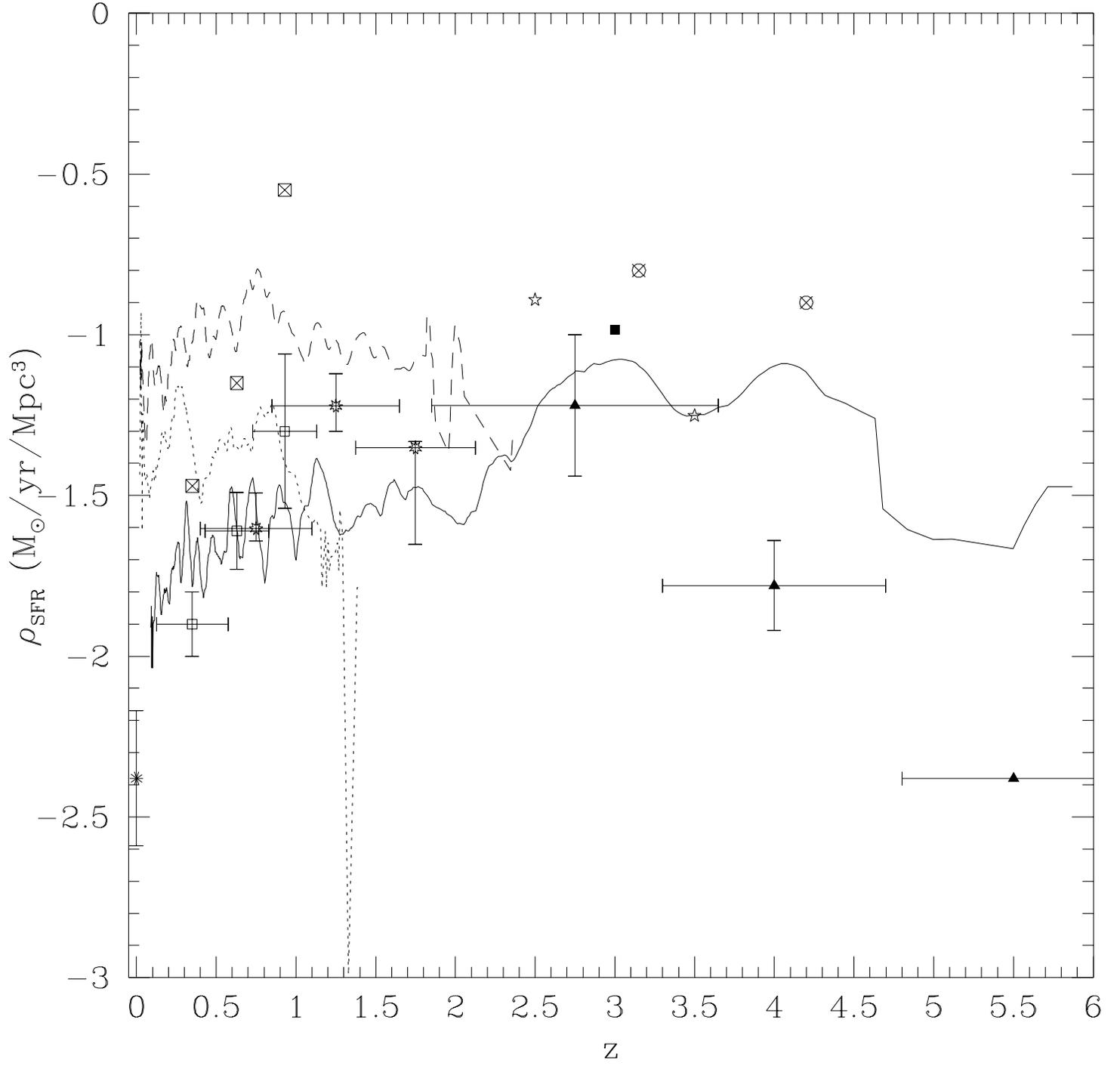
\end{document}